\documentclass[lettersize,journal]{IEEEtran}
\usepackage{amsmath,amsfonts}
\usepackage{algorithm}
\usepackage{array}
\usepackage[caption=false,font=normalsize,labelfont=sf,textfont=sf]{subfig}
\usepackage{textcomp}
\usepackage{stfloats}
\usepackage{url}
\usepackage{verbatim}
\usepackage{cite}

\usepackage{algorithm}
\usepackage{algpseudocode}
\usepackage{tabularx}
\usepackage{booktabs}
\usepackage{multirow}
\usepackage{amssymb,bbding,pifont,makecell,bm}
\newcommand{\cmark}{\ding{51}}%

\usepackage{newfloat}
\usepackage{listings}
\usepackage{graphicx}
\usepackage{xurl}
\usepackage{xparse}

\usepackage{helvet}
\usepackage{courier}
\usepackage{colortbl}  
\usepackage[table,xcdraw]{xcolor}
\usepackage[normalem]{ulem}
\usepackage{ragged2e}

\begin{document}

\title{Wav2code: Restore Clean Speech Representations via Codebook Lookup for Noise-Robust ASR}

\author{Yuchen Hu, Chen Chen, Qiushi Zhu, Eng Siong Chng
\thanks{Manuscript created April 2023, revised Aug 2023, accepted Oct 2023, published Dec 2023.
The computational work for this article was partially performed on resources of the National Supercomputing Centre, Singapore (https://www.nscc.sg).
(Corresponding author: Chen Chen.)}
\thanks{The authors Yuchen Hu, Chen Chen and Eng Siong Chng are with the School of Computer Science and Engineering, Nanyang Technological University (NTU), Singapore 639798 (e-mail: yuchen005@e.ntu.edu.sg; chen1436@e.ntu.edu.sg; aseschng@ntu.edu.sg), and the author Qiushi Zhu is with the National Engineering Research Center of Speech and Language Information Processing (NERC-SLIP), University of Science and Technology of China (USTC), Hefei 230026, China (e-mail:
qszhu@mail.ustc.edu.cn).}}

\markboth{IEEE/ACM TRANSACTIONS ON AUDIO, SPEECH, AND LANGUAGE PROCESSING}%
{Shell \MakeLowercase{\textit{et al.}}: A Sample Article Using IEEEtran.cls for IEEE Journals}


\maketitle

\begin{abstract}
Automatic speech recognition (ASR) has gained remarkable successes thanks to recent advances of deep learning, but it usually degrades significantly under real-world noisy conditions.
Recent works introduce speech enhancement (SE) as front-end to improve speech quality, which is proved effective but may not be optimal for downstream ASR due to speech distortion problem.
Based on that, latest works combine SE and currently popular self-supervised learning (SSL) to alleviate distortion and improve noise robustness.
Despite the effectiveness, the speech distortion caused by conventional SE still cannot be cleared out.
In this paper, we propose a self-supervised framework named Wav2code to implement a feature-level SE with reduced distortions for noise-robust ASR.
First, in pre-training stage the clean speech representations from SSL model are sent to lookup a discrete codebook via nearest-neighbor feature matching, the resulted code sequence are then exploited to reconstruct the original clean representations, in order to store them in codebook as prior.
Second, during finetuning we propose a Transformer-based code predictor to accurately predict clean codes by modeling the global dependency of input noisy representations, which enables discovery and restoration of high-quality clean representations with reduced distortions.
Furthermore, we propose an interactive feature fusion network to combine original noisy and the restored clean representations to consider both fidelity and quality, resulting in more informative features for downstream ASR.
Finally, experiments on both synthetic and real noisy datasets demonstrate that Wav2code can solve the speech distortion and improve ASR performance under various noisy conditions, resulting in stronger robustness.
\end{abstract}


\vspace{-0.05cm}
\section{Introduction}
\IEEEPARstart{T}{he} research of automatic speech recognition (ASR) {has progressed rapidly} during the past few years thanks to the deep learning techniques~\cite{hinton2012deep,graves2014towards,chorowski2015attention,vaswani2017attention}, which {advances the} neural acoustic modeling~\cite{graves2006connectionist,graves2012sequence,chan2016listen,dong2018speech,hu2023gradient}, language modeling~\cite{mikolov2011strategies,irie2016lstm,chen2023hyporadise,chen2023generative,hu2024large,hu2024gentranslate,chen2024its} and large-scale model training~\cite{li2020comparison,lu2020exploring,radford2022robust}.
Compared to conventional ASR approaches like GMM-HMM~\cite{pujol2004comparison}, the power of neural networks has brought significant improvement to ASR performance as well as a simpler end-to-end training pipeline.
However, despite the effectiveness of neural network-based ASR approaches, their performance usually degrades significantly under real-world noisy scenarios~\cite{gong1995speech,li2014overview,krishna2019speech}.

Recently, there are many works focusing on noise-robust speech recognition that explore novel model architectures and training strategies~\cite{chen2022noise,chen2023leveraging,hu2023hearing,hu2023mir,hu2023cross,zhu2024multichannel}. Among them are two successful categories: 1) joint speech enhancement (SE) and ASR systems, 2) self-supervised learning (SSL).
Speech enhancement~\cite{wang2014training,baby2014coupled,chen2023unsupervised,chen2023metric,hu2023unifying,hu2023noise} is originally proposed to reduce additive noise from noisy speech to improve the speech quality for human and machine listening, which can be categorized into time-domain~\cite{luo2018tasnet,pandey2019tcnn,pandey2021dense,lu2022conditional} and frequency-domain~\cite{wang2020complex,yin2020phasen,dang2022dpt} methods.
In particular, time-domain SE aims to reconstruct the target clean speech directly according to the input noisy waveform, while frequency-domain SE receives noisy spectrum as input and {estimates} a mask to obtain the spectrum of target speech, by multiplying the estimated mask with input noisy spectrum.

In view of that, previous works propose to cascade SE and ASR as a joint network~\cite{maas2012recurrent,narayanan2013ideal}, where SE serves as a denoising front-end to benefit downstream ASR.
Regarding the training strategy, several schemes are designed.
First, \cite{narayanan2014joint,gao2015joint,xiao2016deep} proposes end-to-end training of cascaded system using only ASR objective, which can improve ASR results but the front-end SE module is difficult to train well and function without specific training objective.
Another work~\cite{fujimoto2019one} proposes to train SE and ASR modules separately with task-specific objectives.
This strategy usually leads to mismatch between two tasks, as SE is optimized in terms of speech quality like signal-to-noise ratio (SNR), while ASR is optimized in terms of speech intelligibility like cross-entropy and word error rate (WER).
To this end, later works~\cite{wang2016joint,liu2019jointly,ma2021multitask,prasad2021investigation} propose multi-task joint training to optimize SE and ASR modules simultaneously, which results in some improvements.
However, these systems only receives enhanced speech from SE processing as input for downstream ASR, which usually suffers from the speech distortion problem~\cite{iwamoto2022bad}.
In particular, some previous works~\cite{maiti2018large,wang2020bridging} observe that the enhanced speech from SE might not always yield good performance for downstream ASR, as some important ASR-related latent information in original noisy speech are suppressed by SE processing together with the noise, which is often undetected at speech enhancement stage but could be detrimental to the downstream ASR task.
To alleviate this issue, recent works~\cite{hu2022interactive,hu2022dual,iwamoto2022bad} propose to fuse the distorted enhanced speech with original noisy speech to recover some over-suppressed information, which have achieved considerable improvements of ASR performance though still cannot clear out the distortions.

On the other hand, self-supervised learning also shows great potential for noise-robust ASR~\cite{chiu2022self}.
SSL is originally motivated to leverage large amounts of unlabeled data for neural network training.
For instance, TERA~\cite{liu2021tera} performs contextual speech representation learning with BERT~\cite{devlin2018bert} by reconstructing the masked input frames in an utterance.
Similarly, Wav2vec2.0~\cite{baevski2020wav2vec} employs a convolutional neural network (CNN) downsampler to extract local speech features from raw waveform, which are then sent into BERT to perform mask prediction with contrastive loss.
HuBERT~\cite{hsu2021hubert} performs offline clustering to provide labels for mask prediction, and WavLM~\cite{chen2022wavlm} introduces an utterance mixing data augmentation method to improve HuBERT.
Data2vec~\cite{baevski2022data2vec} follows the teacher-student training scheme~\cite{tarvainen2017mean} and adopts continuous contextual representations as targets to preform regression tasks.
Recently, there are increasing SSL works to improve the noise robustness of ASR.
Wav2vec-switch~\cite{wang2022wav2vec} models the noise robustness into contextual speech representations by contrastive learning.
Enhanced Wav2vec2.0 (EW2)~\cite{zhu2022noise} performs mask prediction with noisy speech representations and clean quantized targets, which results in stronger robustness against various noisy conditions than vanilla Wav2vec2.0.

Since both speech enhancement and self-supervised learning can improve the robustness of ASR, latest works~\cite{zhu2022joint,chang2022end} propose to integrate them as an end-to-end network to further improve noise robustness and alleviate the speech distortion caused by SE.
Based on EW2, \cite{zhu2022joint} employs SE as front-end module to produce enhanced speech for SSL, and then performs mask prediction with enhanced speech representations and clean quantized targets, which results in stronger robustness against both noise and speech distortion.
However, this method still only alleviates the speech distortion caused by conventional SE instead of clearing it out, and we also note that the unlabeled clean speech is not fully utilized in their method, \textit{i.e.}, performed as pre-training targets only.

We may gain inspirations from the recently popular vector quantization (VQ)~\cite{jegou2010product,shi2021discretization}, an effective method for discrete representation learning.
VQ has been popular in many generative tasks due to its strong encoding ability in discrete finite space with less mapping uncertainty and higher expressiveness, including speech coding~\cite{jiang2022cross}, audio compression~\cite{zeghidour2021soundstream,defossez2022high}, audio generation~\cite{van2017neural}, image synthesis~\cite{esser2021taming,zhou2022towards,sun2022learning}, etc.
In particular,~\cite{van2017neural} incorporates the idea of VQ into variational autoencoder (VAE) to learn discrete latent representations, which supports high-quality generation of image, video and speech, as well as high quality speaker conversion and unsupervised learning of phonemes.
Furthermore,~\cite{esser2021taming} employs a convolutional VQGAN to learn a codebook of context-rich visual parts, whose composition is subsequently modeled with an autoregressive Transformer architecture, which gains promising performance on high-resolution image synthesis.
Based on that,~\cite{zhou2022towards} introduces a global prediction network for better code composition, which has achieved great success in image super-resolution.
Inspired by them, in this work we leverage VQ in noise-robust ASR to generate clean speech representations {with reduced} distortions from noisy inputs.


Motivated by above observations, in this paper we propose a self-supervised framework based on EW2 named Wav2code, which makes full use of the unlabeled clean speech data as prior to implement {a feature-level SE with reduced} distortions for noise-robust speech recognition.
First, during pre-training we feed the clean waveform into { pre-trained} EW2 from~\cite{zhu2022noise} to obtain SSL representations, which are employed to lookup a discrete codebook via NN feature matching.
The resulted code sequence is then exploited to reconstruct the input clean speech representations.
{ By optimizing such reconstruction task, the codebook would be updated to store clean speech representations as priors.}
Second, in finetuning stage we propose a Transformer-based code predictor to accurately predict clean codes by modeling the global dependency of input noisy representations, which enables discovery and restoration of high-quality clean speech representations.
{Compared to prior works on VQ~\cite{zeghidour2021soundstream,defossez2022high}, our pre-trained clean prior in discrete space and global modeling of input noisy speech representations in Wav2code can effectively remedy the speech distortions that exist in conventional SE.}
Furthermore, considering that the restored clean speech representations from discrete codebook may lose some fidelity (see Fig.~\ref{fig7} (b)), we propose an interactive feature fusion network (IFF-Net) based on our conference paper~\cite{hu2022interactive} to combine original noisy and the restored clean representations to consider both fidelity and quality, resulting in more informative features for downstream ASR.
Finally, experimental results on both the synthetic noisy LibriSpeech data and the real noisy CHiME-4 data demonstrate that our proposed Wav2code achieves consistent ASR improvements under various noisy conditions, which shows stronger noise robustness.
In addition, visualizations show that our Wav2code can learn a good codebook to represent clean speech features, based on that the code predictor can restore high-quality clean representations {with reduced} distortions, and the proposed IFF-Net can achieve good restoration fidelity and quality simultaneously to benefit downstream ASR.

The remainder of this paper is organized as follows.
Section~\ref{sec:method} presents the proposed Wav2code framework including EW2 backbone, codebook learning (pre-training), code prediction (finetuning) and IFF-Net.
Section~\ref{sec:exp_setup} describes the experimental setups, followed by experimental results and analysis in Section~\ref{sec:results_analysis}. 
Section~\ref{sec:conclusion} concludes this work in final.

\section{Methodology}
\label{sec:method}
In this section, we present the proposed Wav2code framework including EW2 backbone, codebook learning in pre-training stage, code prediction in finetuning stage and the IFF-Net.
{The overall training pipeline is depicted by Algorithm~\ref{alg1}.}

\subsection{Backbone: Enhanced Wav2vec2.0 (EW2)}
\label{ssec:backbone_ew2}
\begin{figure}[t]
  \centering
  \includegraphics[width=0.44\textwidth]{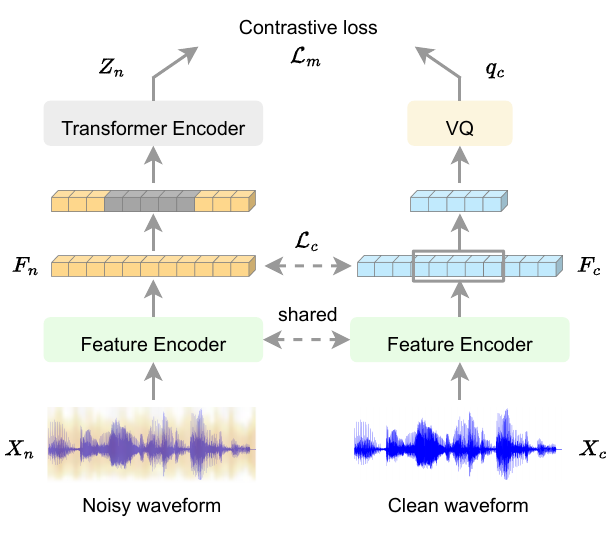}
  \vspace{-0.2cm}
  \caption{Illustration of Enhanced Wav2vec2.0.
  It is pre-trained with contrastive loss between masked noisy speech representations $Z_n$ and quantized clean targets $q_c$, and then finetuned with a linear output layer and CTC loss~\cite{graves2006connectionist}.
  }
  \vspace{-0.2cm}
  \label{fig1}
\end{figure}

We employ the Enhanced Wav2vec2.0 (EW2)~\cite{zhu2022noise} as our backbone, which is an extension of vanilla Wav2vec2.0~\cite{baevski2020wav2vec}. 
As shown in Fig.~\ref{fig1}, EW2 consists of a feature encoder $f$, a Transformer encoder $g$ and a vector quantization (VQ) module.
First, the feature encoder comprises seven CNN layers to downsample the input noisy and clean raw waveforms, generating two speech features $F_n, F_c \in \mathbb{R}^{T\times D}$, where $T$ is number of frames and $D$ is embedding dimension,
\begin{equation}
  F_n = f(X_n);\quad
  F_c = f(X_c),
  \label{eq1}
\end{equation}

Then, a certain proportion of frames in noisy features $F_n$ are masked by replaced with a learnable mask embedding, which is sent into the Transformer encoder to learn a high-level contextualized representation $Z_n \in \mathbb{R}^{T\times D}$,
\begin{equation}
  Z_n = g(F_n),
  \label{eq2}
\end{equation}
Meanwhile, the parallel clean features $F_c$ are discretized into $q_c$ using a VQ module, which provides clean targets for subsequent contrastive pre-training,
\begin{equation}
  q_c = VQ(F_c),
  \label{eq3}
\end{equation}
where $q_c \in \mathbb{R}^{T'\times D}$, and $T'$ is the number of masked frames.
The motivation of using clean features as targets is to encourage the model to predict clean speech representations from noisy speech inputs, and thus enhances its noise robustness.
In particular, the implementation details of EW2 follow that in vanilla Wav2vec2.0, including model settings, VQ configurations and the training details.
The final pre-training objective can be formulated as
\begin{equation}
  \mathcal{L} = \mathcal{L}_{m}+\alpha \mathcal{L}_{d} + \beta \mathcal{L}_{f} + \gamma \mathcal{L}_{c},
  \label{eq4}
\end{equation}
where
\begin{align}
  \mathcal{L}_{m} &= -\mathbb{E}_t\left(\log \frac{\exp({\rm sim}(Z_{n_t},q_{c_t})/\kappa)}{\sum_{\tilde{q} {\sim} Q_{t}}\exp({\rm sim}(Z_{n_t},\tilde{q})/\kappa)}\right),
  \label{eq5} \\
  \mathcal{L}_{c} &= \mathbb{E}_t\left(\left\| F_{n_{t}}-F_{c_{t}} \right\|_2^2\right),
  \label{eq6}
\end{align}
the subscript $t$ indicates frame index.
$\mathcal{L}_m$ is the contrastive loss that enables the model to distinguish between the true quantized clean feature $q_{c_t}$ and a set of $K+1$ quantized candidature features $\tilde{q} \in Q_t$, where $Q_t$ contains $q_{c_t}$ and $K$ distractors sampled from other frames.
$\rm sim(\cdot,\cdot)$ denotes cosine similarity and $\kappa$ is a temperature.
$\mathcal{L}_d$ is a diversity loss to encourage exhausted use of all the codebook entries, and $\mathcal{L}_f$ is a $\ell_2$ penalty over the outputs of feature encoder, which directly follows vanilla Wav2vec2.0~\cite{baevski2020wav2vec}.
In addition, $\mathcal{L}_c$ is $\ell_2$ loss to ensure the consistency between noisy and clean speech features, by minimizing their Euclidean distance.
The final loss function $\mathcal{L}$ is a weighted summation of above four loss terms with weight hyper-parameters $\alpha$, $\beta$ and $\gamma$.

\begin{figure*}[t]
  \centering
  \includegraphics[width=1.0\linewidth]{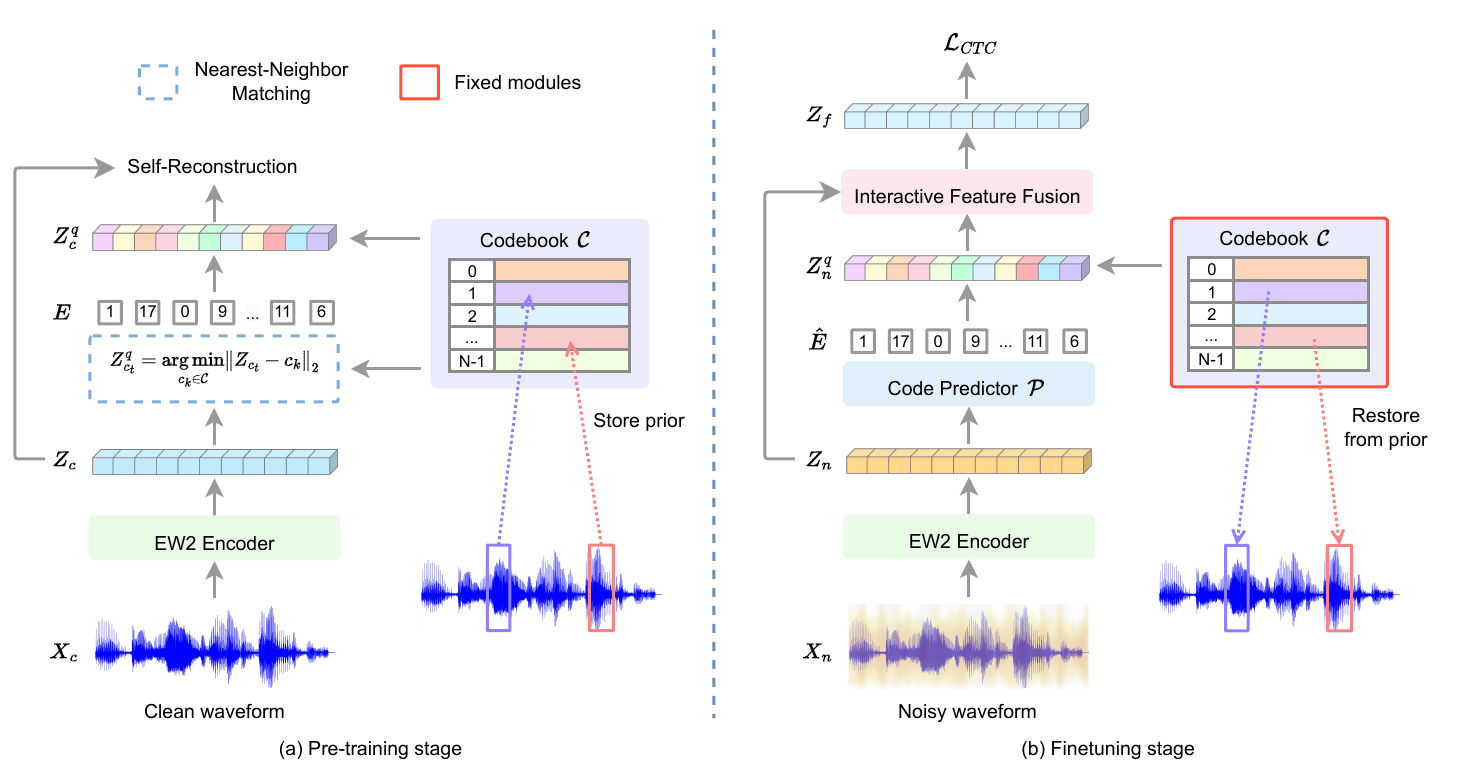}
  \vspace{-0.5cm}
  \caption{Illustration of the proposed Wav2code framework.
  (a) Pre-training stage: codebook learning via nearest-neighbor matching to store clean speech prior.
  (b) Finetuning stage: code prediction to restore clean speech representations from prior for downstream ASR.
  Solid arrows indicate direct data flow, and dashed arrows indicate mapping relationship.
  }
  \vspace{-0.2cm}
  \label{fig2}
\end{figure*}

\subsection{{Wav2code Pre-training}: Codebook Learning}
\label{ssec:pre_train}
To reduce the uncertainty of noisy-to-clean speech representation mapping in conventional SE and recover the distorted information in enhanced speech, we first pre-train Wav2code to learn a codebook with rich prior of clean speech representations, which strengthen the expressiveness of entire system and thus support more robust {feature-level} SE in finetuning.

As illustrated in Fig.~\ref{fig2}(a), the clean raw waveform $X_c$ is first embedded as contextualized speech representation $Z_c \in \mathbb{R}^{T\times D}$ by a pre-trained EW2 encoder, which contains a feature encoder and a Transformer encoder as shown in Fig.~\ref{fig1}, and the training details is explained in Section~\ref{ssec:exp_config},
\begin{equation}
  Z_c = EW2(X_c),
  \label{eq7}
\end{equation}
Following the popular VQ-VAE~\cite{van2017neural} and VQGAN~\cite{esser2021taming}, we replace each frame in clean representation $Z_c$ (\textit{i.e.}, $Z_{c_t}$, $t\in\{0,1,2,...,T-1\}$ is frame index) with the nearest entry in the discrete codebook $\mathcal{C}=\{c_k\in\mathbb{R}^D\}_{k=0}^{N-1}$, in order to obtain the quantized clean speech feature $Z_c^q \in \mathbb{R}^{T\times D}$ as well as the corresponding codebook entry ID sequence $E=\{e_t\in\{0,1,2,...,N-1\}\}_{t=0}^{T-1}$,
\begin{align}
  Z_{c_t}^q = \underset{c_k\in\mathcal{C}}{\arg\min} \left\|Z_{c_t}-c_k\right\|_2,
  \label{eq8} \\
  e_t = \underset{k}{\arg\min}\left\|Z_{c_t}-c_k\right\|_2,
  \label{eq9}
\end{align}
{The quantized clean feature $Z_c^q$ actually represents a sequence of exemplars in codebook space.
Then, we perform to reconstruct the input clean speech $Z_c$ from this code sequence $Z_c^q$ by Eq.(\ref{eq10}), which is similar to k-means clustering.
As a result, the codebook $\mathcal{C}$ will be updated to store priors from clean speech representations, resulting in deep correlations between clean speech space and codebook space.
In addition, $E$ is a sequence of codebook entry IDs that corresponds to each frame in $Z_c^q$, \textit{i.e.}, $e_t = k$ if $Z_{c_t}^q = c_k$, where $k\in\{0,1,2...,N-1\}$}.

Inspired by~\cite{van2017neural,esser2021taming}, we build a bi-directional pre-training loss to reconstruct the clean representation $Z_c$ from $Z_c^q$,
\begin{equation}
  \mathcal{L}_{pt} = \left\| \text{sg}(Z_c) - Z_c^q \right\|_2^2 + \beta \left\| Z_c - \text{sg}(Z_c^q) \right\|_2^2,
  \label{eq10}
\end{equation}
where $\text{sg}(\cdot)$ denotes the stop-gradient operator.
The first term is reconstruction loss to train the codebook $\mathcal{C}$, in order to store the clean speech prior.
The second term is a commitment loss~\cite{van2017neural} to align the optimization of front-end encoder and codebook.
We find that jointly training front-end EW2 encoder and codebook benefits the prior learning, resulting in better codebook for subsequent finetuning process.

The weight $\beta$ is designed for trade-off between the update rates of EW2 encoder and codebook, which is empirically set to 0.25.
The code dimension and embedding dimension $D$ is set to 768.
For the number of codebook entries $N$, we empirically set it to 1024, which is sufficient to restore clean speech representations.
Further increasing it can only yield limited improvement, as the redundancy in codebook entries could cause ambiguous code prediction in finetuning stage, where the experimental analysis is presented in Section~\ref{ssec:eval_codebook_codepred}.

\subsection{{Wav2code Finetuning}: Code Prediction}
\label{ssec:finetune}
In finetuning stage, we aim to restore clean speech representations from the pre-trained codebook prior according to the noisy inputs, and thus benefit the downstream ASR.
However, due to the noise corruption in input noisy representations, the nearest-neighbor (NN) matching in Eq.~\ref{eq8} usually fails to find the correct codebook entry for clean representation restoration.
As shown in Fig.~\ref{fig5}(b), the uncertain and diverse noise corruption deviate the noisy speech representations from the correct codebook entry, which thus falls in nearby clusters and leads to sub-optimal restoration results (see Fig.~\ref{fig6}(c)).
To this end, we introduce a Transformer-based code predictor to model the global contextual dependencies of input noisy sequence for accurate code prediction.

Similar to pre-training stage, EW2 encoder is employed to process the input noisy raw waveform and generate a noisy representation $Z_n\in\mathbb{R}^{T\times D}$.
Then, we design a Transformer-based code predictor $\mathcal{P}$ to model the global dependencies of $Z_n$.
In particular, $\mathcal{P}$ first contains $M$ Transformer encoder blocks, where the $m$-th block is formulated as,
\begin{align}
  H_m &= LN(X_m + M\hspace{-0.02cm}H\hspace{-0.02cm}A(X_m, X_m, X_m)),
  \label{eq11} \\
  X_{m+1} &=LN(H_m + F\hspace{-0.02cm}F\hspace{-0.02cm}N(H_m)),
  \label{eq12}
\end{align}
where $X_0 = \text{Project}(Z_n) \in\mathbb{R}^{T\times D_p}$ and $D_p<D$ is to enable efficient Transformer, ``LN'' denotes layer normalization~\cite{ba2016layer}, ``MHA'' denotes multi-head attention~\cite{vaswani2017attention}.
After Transformer encoder blocks, there is a linear projection to map $X_M\in\mathbb{R}^{T\times D_p}$ to $\mathbb{R}^{T\times N}$, followed by a softmax layer to generate code prediction outputs $\hat{P}\in\mathbb{R}^{T\times N}$, where $\hat{p_t}\in\mathbb{R}^N$ is a probability distribution over $N$ codebook entries.

Meanwhile, in order to further retrieve the corresponding codebook entries for each frame according to the predicted probability distribution, we employ the gumbel softmax layer~\cite{jang2016categorical} to select discrete codebook entries in a fully differentiable way, which generates a predicted codebook entry ID sequence $\hat{E}\in\mathbb{N}^{T\times1}$, where $\hat{e_t}\in\{0,1,2,...,N-1\}$.
It then retrieves $T$ respective entries from the pre-trained codebook to form the quantized representation $Z_n^q\in\mathbb{R}^{T\times D}$, which restores high-quality clean speech feature to benefit downstream ASR.
{Therefore, it produces a noisy-to-clean transformation like conventional SE but on representations instead of raw speech.}

Finally, the restored representation $Z_n^q$ and original noisy representation $Z_n$ {are combined (see Sec.~\ref{ssec:iffnet}) and sent for downstream ASR} with a linear layer and CTC loss~\cite{graves2006connectionist} $\mathcal{L}_{CTC}$.

For training process, we first load the pre-trained EW2 backbone from Section~\ref{ssec:backbone_ew2} and codebook $\mathcal{C}$ from pre-training stage in Section~\ref{ssec:pre_train}.
Codebook is fixed during finetuning to protect the clean speech prior.
We exploit two training objectives in this stage:
1) cross-entropy loss $\mathcal{L}_{pred}$ for code prediction supervision, and 
2) $\ell_2$ loss $\mathcal{L}_{res}$ for clean feature restoration supervision,
\begin{equation}
  \mathcal{L}_{pred} = \mathbb{E}_t\left(-e_t\log(\hat{e_t})\right);\quad
  \mathcal{L}_{res} = \left\|Z_n^q - sg(Z_c)\right\|_2^2,
  \label{eq13}
\end{equation}
where the ground-truth of code $E$ is obtained from pre-training stage in Section~\ref{ssec:pre_train}, and $Z_c$ is the ground-truth clean representation to supervise the restoration.
The final objective of finetuning stage is,
\begin{equation}
  \mathcal{L}_{ft} = \mathcal{L}_{CTC} + \lambda_{pred} \cdot \mathcal{L}_{pred} + \lambda_{res} \cdot \mathcal{L}_{res},
  \label{eq14}
\end{equation}
where $\lambda_{pred}$ and $\lambda_{res}$ are both set to 0.1 in our system.

\subsection{Interactive Feature Fusion Network (IFF-Net)}
\label{ssec:iffnet}

\begin{figure}[t]
  \centering
  \includegraphics[width=0.44\textwidth]{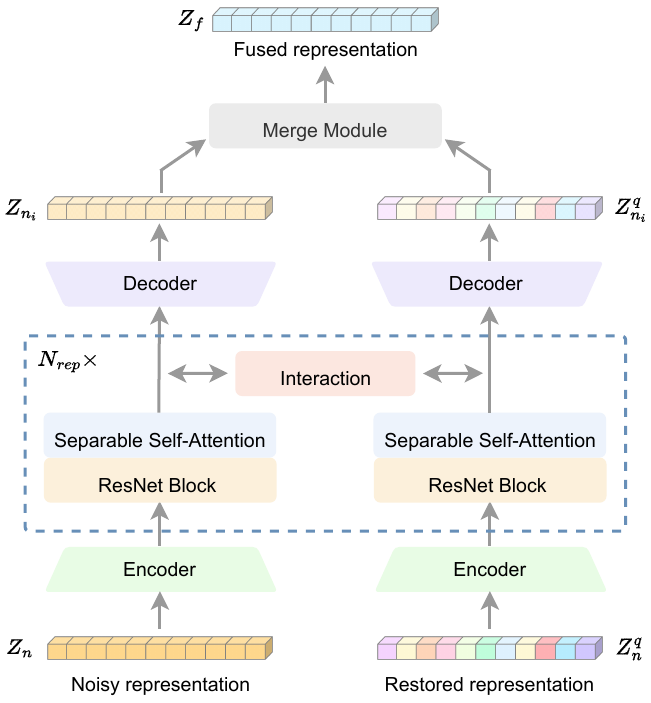}
  \vspace{-0.1cm}
  \caption{Illustration of proposed interaction feature fusion network (IFF-Net).
  Encoder and Decoder are used to compress and recover the number of feature channels like bottleneck.
  ResNet block is employed to capture local context, and Separable Self-Attention (SSA) module is exploited to model global dependencies.
  Interaction module is designed to interact between two branch of features, which are finally fused by Merge module to generate output.
  }
  \label{fig3}
  \vspace{-0.1cm}
\end{figure}

Since the quantized speech representation $Z_n^q$ is restored from discrete codebook, there exists some loss of fidelity (see Fig.~\ref{fig7}(b)), though its quality has been improved significantly.
On the other hand, the fidelity of original noisy representation $Z_n$ is unaffected, while its quality suffers from noise corruption.
Therefore, some fusion strategy between these two features may result in more informative speech representations to benefit downstream ASR.
To this end, we propose an interaction feature fusion network (IFF-Net) to combine them, in order to purse both fidelity and quality in restoration.

In particular, there is first a bottleneck design with a pair of CNN encoder and decoder to enable efficient computation.
The bottleneck features in-between are all of tensor shape $\mathbb{R}^{T\times D'}$, where $D'$ is empirically set to 128 in our system.

To capture both local and global dependencies of the two input representations for further interaction, we leverage a ResNet block and a proposed Separable Self-Attention (SSA)~{\cite{hu2022interactive}} module for sequence modeling.
The ResNet block consists of one 1-D convolutional layer with kernel size of 3, stride of 1 and channel number of $D'$, as well as a residual connection to extract deep local features from inputs.
These features are then fed into a SSA module to capture global dependencies along both temporal and embedding dimensions.
Take the branch of noisy representation for example, denote the SSA module input as $H_n\in\mathbb{R}^{T\times D'}$, the temporal- and embedding-wise self-attention are respectively formulated as,
\begin{align}
  H_n^{temp} &= H_n + \text{Softmax}(H_n\cdot (H_n)^T/\sqrt{D'})\cdot H_n,
  \label{eq15} \\
  H_n^{embd} &= H_n + H_n \cdot \text{Softmax}((H_n)^T\cdot H_n/\sqrt{T}),
  \label{eq16}
\end{align}
Then SSA output is obtained by combining the global features,
\begin{equation}
  S_n = \text{Conv}(H_n \Vert H_n^{temp} \Vert H_n^{embd}),
  \label{eq17}
\end{equation}
where $\Vert$ denotes embedding-wise feature concatenation, and ``Conv'' denotes 1$\times$1 convolution layer.

After sequence modeling, the two branch of features {interact} to augment each other,
\begin{align}
  S_{n_i} &= S_n + \sigma(\text{Conv}(S_n\Vert S_n^q)) \otimes S_n^q,
  \label{eq18} \\
  S_{n_i}^q &= S_n^q + \sigma(\text{Conv}(S_n\Vert S_n^q)) \otimes S_n,
  \label{eq19}
\end{align}
where $\sigma$ denotes the Sigmoid activation function, $\otimes$ denotes element-wise multiplication.
This Interaction module learns a mask to weight the two features to augment each other.

The bottleneck features are processed by ResNet block, SSA module and Interaction module for $N_{rep}$ repeated times before mapped back to $D$-channel feature space by decoder, where we set $N_{rep}=4$ in our system.
Then, the decoder outputs $Z_{n_i}$ and $Z_{n_i}^q$ are finally sent into a Merge module for fusion.
Specifically, this module learns a element-wise mask that acts as a gate to control which parts of final representations focus on fidelity and which parts focus on quality, and the final output is obtained as weighted sum,
\begin{align}
  M &= \sigma(\text{Self-Attention}(\text{Conv}(Z_{n_i}\Vert Z_{n_i}^q))),
  \label{eq20} \\
  Z_f &= Z_{n_i} \otimes M + Z_{n_i}^q \otimes (1-M),
  \label{eq21}
\end{align}

As a result, the proposed IFF-Net can extract deep interactive features of two branches and finally generate a fused representation to pursue both fidelity and quality.

\begin{algorithm}[t]

\caption{Wav2code training pipeline in three stages.}
\label{alg1}
\small
\begin{algorithmic}[1]
    \Require EW2 encoder, codebook, code predictor, IFF-Net, etc.
    \State Conduct EW2 backbone pre-training following \S\ref{ssec:backbone_ew2};
    \State Load pre-trained weights of EW2 encoder from stage 1 and perform Wav2code pre-training following \S\ref{ssec:pre_train};
    \State Load pre-trained weights of EW2 encoder and codebook from stage 2, conduct Wav2code finetuning following \S\ref{ssec:finetune};
\end{algorithmic}
\normalsize
\end{algorithm}

\section{Experimental Setup}
\label{sec:exp_setup}
In this section, we introduce the experimental setup in detail, including datasets for system evaluation, experimental configurations for EW2 backbone pre-training, Wav2code pre-training and Wav2code finetuning.

\vspace{-0.1cm}
\subsection{Datasets}
\label{ssec:datasets}
\textbf{LibriSpeech{-FreeSound}~\cite{prasad2021investigation}:} For fair comparison with existing baseline approaches, we use the same dataset as that in~\cite{prasad2021investigation}.
Particularly, we utilize LibriSpeech~\cite{panayotov2015librispeech} \textit{train-clean-100} and \textit{train-clean-360} subsets as clean training data, and the \textit{dev-clean} subset as { validation} data.
To simulate noisy speech data to augment model training and validation, we randomly select noise samples from FreeSound~\cite{prasad2021investigation} dataset and mix them with the LibriSpeech clean samples, at a randomly selected SNR from $\{0,5,10,15,20,25\}$ dB.
The noisy test set is available online\footnote{https://github.com/archiki/Robust-E2E-ASR}, which contains 4200 noisy speech samples simulated by selecting 120 clean speech samples from LibriSpeech \textit{test-clean} subset and mixing each of them with 7 different noise types at 5 different SNRs from $\{0,5,10,15,20\}$ dB.
The FreeSound noise data is sampled at 16 kHz, with seven noise types that are divided into two categories, \textit{i.e.}, type-A and type-B.
The type-A noise is stationary, including \textit{`Traffic', `Metro', `Car'}, while the type-B noise is relatively non-stationary, including \textit{`Babble', `Airport/Station', `AC/Vacuum', `Cafe'}.
There are 10 and 8 different audio samples for each noise type in training and test sets, respectively, and the total duration of noise data is around 2 hours.

\textbf{CHiME-4~\cite{vincent2016chime4}:} To further validate our proposed approach on real noisy data, we also conduct experiments on CHiME-4 challenge data\footnote{http://spandh.dcs.shef.ac.uk/chime\_challenge/CHiME4/index.html}~\cite{vincent2016chime4}.
The CHiME-4 dataset is collected by asking volunteers to read the text of Wall Street Journal (WSJ0) corpus~\cite{paul1992design} and using a six-channel distant microphone array as well as a close-talk microphone to record data.
There are two types of noisy speech data, \textit{i.e.}, real and simulated.
The real noisy speech is recorded in four different environments including bus, cafe, pedestrian area and street junction, and the simulated noisy data is synthesized by mixing clean speech from WSJ0 training set (si\_tr\_s) and the noise recorded in above four environments.
The entire dataset is divided into three splits, \textit{i.e.}, training, { validation} and test.
The training split contains 1600 real noisy and 7138 simulated noisy utterances, the { validation} split contains 1640 real noisy and 1640 simulated noisy utterances, and the test split contains 1320 real noisy and 1320 simulated noisy utterances.
Follow prior work~\cite{zhu2022joint}, we use the six-channel data for training, and the one-channel real noisy data for validation and testing.

\begin{table*}[t]
\caption{The performance comparison of different approaches on type-A noisy test sets at different SNR levels.
For all the baselines, ``Pre-train'' and ``Finetine'' respectively denotes the {data type} for self-supervised pre-training and finetuning.
For the proposed Wav2code, ``Pre-train'' denotes the {data type} for EW2 backbone pre-training and ``Finetine'' denotes that for Wav2code finetuning, note that Wav2code pre-training requires clean data only.
``None'' denotes {no SSL pre-training}.
``Avg'' denotes the averaged WER results on all type-A noise conditions, and ``Clean'' denotes the WER results on clean test set.}
\label{table1}
\centering
\resizebox{1.0\textwidth}{!}{
	\begin{tabular}{l|c|c|ccccccccccccccccc}
		\hline
		\multirow{3}{*}{\textbf{Method}}     & \multirow{3}{*}{\textbf{Pre-train}} & \multirow{3}{*}{\textbf{Finetune}} & \multicolumn{17}{c}{\textbf{WER under SNR (dB)}}\\ \cline{4-20} 
		&  &  & \multicolumn{5}{c|}{\textbf{Traffic}} & \multicolumn{5}{c|}{\textbf{Metro}} & \multicolumn{5}{c|}{\textbf{Car}} & \multicolumn{1}{c|}{\multirow{2}{*}{\textbf{Avg}}} & \multirow{2}{*}{\textbf{Clean}} \\ \cline{4-18}
		& & & \textbf{0} & \textbf{5} & \textbf{10} & \textbf{15} & \multicolumn{1}{c|}{\textbf{20}} & \textbf{0} & \textbf{5} & \textbf{10} & \textbf{15} & \multicolumn{1}{c|}{\textbf{20}} & \textbf{0} & \textbf{5} & \textbf{10} & \textbf{15} & \multicolumn{1}{c|}{\textbf{20}} & \multicolumn{1}{c|}{} & \\ 
		\hline
		Baseline~\cite{prasad2021investigation} & None & Clean & 72.4 & 62.5 & 50.2 & 41.0 & \multicolumn{1}{c|}{33.6} & 68.4 & 54.4 & 46.4 & 34.9 & \multicolumn{1}{c|}{27.6} & 35.0 & 28.1 & 24.3 & 21.7 & \multicolumn{1}{c|}{16.7} & \multicolumn{1}{c|}{41.1} & \textbf{10.3} \\ 
		DEMUCS~\cite{prasad2021investigation} & {Noisy} & {Noisy} & 38.2 & 30.3 & 25.3 & 20.6 & \multicolumn{1}{c|}{17.9} & 35.6 & 24.9 & 22.6 & 17.1 & \multicolumn{1}{c|}{15.9} & 20.5 & 18.1 & 14.6 & 13.8 & \multicolumn{1}{c|}{13.1} & \multicolumn{1}{c|}{21.9} & 10.9 \\
        AvT~\cite{prasad2021investigation} & None & {Noisy} & 40.7 & 32.5 & 26.3 & 21.4 & \multicolumn{1}{c|}{18.5} & 36.1 & 26.5 & 22.6 & 18.4 & \multicolumn{1}{c|}{17.8} & 21.8 & 18.9 & 16.8 & 16.0 & \multicolumn{1}{c|}{15.3} & \multicolumn{1}{c|}{23.3} & 13.1 \\
        \hline
        \multirow{4}{*}{Wav2vec2.0~\cite{baevski2020wav2vec}} & None & Clean & 71.0 & 57.8 & 48.6 & 40.5 & \multicolumn{1}{c|}{33.3} & 66.5 & 54.4 & 48.6 & 35.2 & \multicolumn{1}{c|}{28.2} & 34.1 & 27.1 & 22.8 & 19.2 & \multicolumn{1}{c|}{14.8} & \multicolumn{1}{c|}{40.1} & 11.0 \\
        & None & {Noisy} & 52.3 & 44.8 & 38.9 & 34.6 & \multicolumn{1}{c|}{31.2} & 49.9 & 41.1 & 36.3 & 31.6 & \multicolumn{1}{c|}{30.0} & 34.7 & 30.3 & 29.5 & 28.4 & \multicolumn{1}{c|}{26.8} & \multicolumn{1}{c|}{36.0} & 25.0 \\
         & Clean & {Noisy} & 42.8 & 34.2 & 26.7 & 23.0 & \multicolumn{1}{c|}{19.4} & 39.2 & 31.0 & 25.9 & 21.4 & \multicolumn{1}{c|}{19.7} & 23.0 & 19.0 & 17.6 & 16.2 & \multicolumn{1}{c|}{15.4} & \multicolumn{1}{c|}{25.0} & 14.0 \\
        & {Noisy} & {Noisy} & 34.6 & 27.9 & 23.6 & 18.9 & \multicolumn{1}{c|}{17.6} & 32.4 & 24.8 & 20.7 & 17.4 & \multicolumn{1}{c|}{17.1} & 19.4 & 17.0 & 15.5 & 14.7 & \multicolumn{1}{c|}{14.6} & \multicolumn{1}{c|}{21.1} & 13.5 \\
        \hline
        EW2~\cite{zhu2022noise} & {Noisy} & {Noisy} & 31.0 & 22.3 & 20.0 & 16.5 & \multicolumn{1}{c|}{14.9} & 29.0 & 21.9 & 18.1 & 15.8 & \multicolumn{1}{c|}{14.4} & 17.6 & 15.6 & 14.4 & 13.7 & \multicolumn{1}{c|}{13.1} & \multicolumn{1}{c|}{18.6} & 12.3 \\
        \hline
		\multirow{2}{*}{EW2 + SEW2~\cite{zhu2022joint}} & None & {Noisy} & 37.3 & 29.9 & 25.7 & 22.3 & \multicolumn{1}{c|}{19.8} & 36.7 & 27.5 & 24.3 & 20.5 & \multicolumn{1}{c|}{19.2} & 22.2 & 18.9 & 18.3 & 17.6 & \multicolumn{1}{c|}{16.8} & \multicolumn{1}{c|}{23.8} & 15.6 \\
		& {Noisy} & {Noisy} & 28.6 & 21.1 & 18.9 & 15.7 & \multicolumn{1}{c|}{14.5} & 27.5 & 20.9 & 17.4 & 15.3 & \multicolumn{1}{c|}{14.1} & 16.3 & 14.8 & 13.8 & 13.4 & \multicolumn{1}{c|}{13.0} & \multicolumn{1}{c|}{17.7} & 12.2 \\
		\hline
		\multirow{2}{*}{Wav2code (ours)} & None & {Noisy} & 34.6 & 27.6 & 23.8 & 20.7 & \multicolumn{1}{c|}{18.6} & 34.4 & 25.7 & 22.7 & 19.2 & \multicolumn{1}{c|}{18.3} & 20.4 & 17.8 & 16.9 & 16.5 & \multicolumn{1}{c|}{16.2} & \multicolumn{1}{c|}{22.2} & 14.7 \\
		& {Noisy} & {Noisy} & \textbf{26.9} & \textbf{20.1} & \textbf{18.2} & \textbf{15.2} & \multicolumn{1}{c|}{\textbf{14.1}} & \textbf{26.5} & \textbf{20.2} & \textbf{16.8} & \textbf{14.9} & \multicolumn{1}{c|}{\textbf{13.8}} & \textbf{15.1} & \textbf{13.8} & \textbf{13.1} & \textbf{12.9} & \multicolumn{1}{c|}{\textbf{12.6}} & \multicolumn{1}{c|}{\textbf{16.9}} & 11.4 \\
		\hline
	\end{tabular}
}
\vspace{-0.2cm}
\end{table*}

\vspace{-0.15cm}
\subsection{Experimental Configurations}
\label{ssec:exp_config}

\textbf{EW2 Backbone Pre-training on LibriSpeech:} We build two settings for EW2 backbone pre-training~\cite{zhu2022joint}, \textit{i.e.}, 100 hours and 460 hours.
For 100-hour setting, we pre-train EW2 backbone on \textit{train-clean-100} subset of LibriSpeech and its noise version with FreeSound augmentation.
For model architecture, the feature encoder contains seven convolutional layers with channel number all of 512, the strides of (5, 2, 2, 2, 2, 2, 2) and the kernel sizes of (10, 3, 3, 3, 3, 2, 2).
As a result, the frame shift of feature encoder output is 20 ms and the receptive field is 25 ms.
The following Transformer encoder consists of 12 layers comprising a self-attention module and a feed-forward network (FFN) module.
The dimension of self-attention module is 768, the number of attention heads is 12, and the inner dimension of FFN is 2048.
For VQ module, we follow the same configurations in prior work~\cite{zhu2022joint}.
The total number of parameters in EW2 backbone is around 95 M.
For masking process, we sample from all the frames at a probability of $p=0.065$ as starting points and then mask the subsequent $M=10$ frames after them.
In calculating contrastive loss by Eq.~\ref{eq5}, the number of distractors $K$ is set to 100, and the temperature $\kappa$ is set to 0.1.
The weighting parameters $\alpha$, $\beta$ and $\gamma$ are set to 0.1, 10 and 1, respectively.
We pre-train EW2 for 100k steps using Adam optimizer~\cite{kingma2014adam}. During the first 20\% of all training steps, the learning rate warms up to $5e{-4}$ and then decays polynomially.

For 460-hour setting, all the configurations are kept same as before except that we pre-train EW2 backbone on a combination of \textit{train-clean-100} and \textit{train-clean-360} subsets of LibriSpeech and its noise version with FreeSound augmentation. 

\textbf{Wav2code Pre-training on LibriSpeech:} The \textit{train-clean-100} subset of LibriSpeech is utilized for Wav2code pre-training, and noise data is not needed in this stage.
For model architecture, the EW2 backbone is initialized from previous pre-training stage with same configurations.
For discrete codebook, the embedding dimension $D$ follows that in EW2, \textit{i.e.}, 768, and the number of codebook entries $N$ is 1024.
We pre-train Wav2code for 50k steps using Adam optimizer. During the first 20\% of all training steps, the learning rate warms up to $5e{-4}$ and then decays polynomially.

\textbf{Wav2code Finetuning on LibriSpeech:}
During finetuning, we utilize both clean and noisy versions of \textit{train-clean-100} subset of LibriSpeech.
The EW2 backbone is initialized from first pre-training stage, and the codebook is initialized from second pre-training stage.
For model architecture, the number of Transformer blocks $M$ in code predictor is set to 4, and the embedding dimension $D_p$ is set to 256.
The total number of parameters in Wav2code is around 104 M.
The Wav2code is finetuned for 50k steps using Adam optimizer.
We use a data augmentation method similar to SpecAugment~\cite{park2019specaugment}, where a time mask same as that in EW2 pre-training stage is employed, as well as a frequency mask at probability of 0.05 and 32 subsequent channels.
For CTC loss, the modeling unit contains 30 characters, including 26 letters and 4 special symbols, and no language model (LM) is used during inference.

\textbf{Training on CHiME-4}: We employ similar experimental configurations on CHiME-4 dataset as those on LibriSpeech dataset, where the differences are explained as follows.
Due to the limited size of CHiME-4 dataset, we utilize all the channels of data except the second microphone channel for model training.
The EW2 pre-training is continued from the public pre-trained Wav2vec2.0 model\footnote{https://dl.fbaipublicfiles.com/fairseq/wav2vec/wav2vec\_small.pt} for 50k steps, following that the Wav2code is pre-trained and finetuned both for 20k steps.
The public pre-trained model is pre-trained on 960 hours data of LibriSpeech with around 95 M parameters.
Following previous works on CHiME-4 dataset, we train a Transformer-based LM with vocabulary of 65k using the text from WSJ corpus, where the implementation details are same as~\cite{zhu2022joint}.

\begin{table*}[t]
\caption{The performance comparison of different approaches on type-B noisy test sets at different SNR levels.
For all the baselines, ``Pre-train'' and ``Finetine'' respectively denotes the {data type} for self-supervised pre-training and finetuning.
For the proposed Wav2code, ``Pre-train'' denotes the {data type} data for EW2 backbone pre-training and ``Finetine'' denotes that for Wav2code finetuning, note that Wav2code pre-training requires clean data only.
``None'' denotes {no SSL pre-training}.
``Avg'' denotes the averaged WER results on all type-B noise conditions, and ``Clean'' denotes the WER results on clean test set.}
\label{table2}
\centering
\resizebox{1.0\textwidth}{!}{
	\begin{tabular}{l|c|c|ccccccccccccccccc}
		\hline
		\multirow{3}{*}{\textbf{Method}}     & \multirow{3}{*}{\textbf{Pre-train}} & \multirow{3}{*}{\textbf{Finetune}} & \multicolumn{17}{c}{\textbf{WER under SNR (dB)}}\\ \cline{4-20} 
		&  &  & \multicolumn{4}{c|}{\textbf{Babble}} & \multicolumn{4}{c|}{\textbf{Airport/Station}} & \multicolumn{4}{c|}{\textbf{AC/Vacuum}} & \multicolumn{4}{c|}{\textbf{Cafe}} & \multirow{2}{*}{\textbf{Avg}} \\ \cline{4-19}
		& & & \textbf{5} & \textbf{10} & \textbf{15} & \multicolumn{1}{c|}{\textbf{20}} & \textbf{5} & \textbf{10} & \textbf{15} & \multicolumn{1}{c|}{\textbf{20}} & \textbf{5} & \textbf{10} & \textbf{15} & \multicolumn{1}{c|}{\textbf{20}} & \textbf{5} & \textbf{10} & \textbf{15} & \multicolumn{1}{c|}{\textbf{20}} & \\ 
		\hline
		Baseline~\cite{prasad2021investigation} & None & Clean & 98.3 & 91.3 & 79.7 & \multicolumn{1}{c|}{65.0} & 84.1 & 73.7 & 60.6 & \multicolumn{1}{c|}{50.0} & 83.1 & 71.5 & 59.5 & \multicolumn{1}{c|}{45.8} & 72.7 & 59.5 & 44.3 & \multicolumn{1}{c|}{33.4} & 67.0 \\
		DEMUCS~\cite{prasad2021investigation} & {Noisy} & {Noisy} & 58.0 & 41.8 & 32.3 & \multicolumn{1}{c|}{25.4} & 45.5 & 33.7 & 25.6 & \multicolumn{1}{c|}{21.5} & 45.4 & 34.2 & 28.1 & \multicolumn{1}{c|}{22.8} & 31.6 & 27.4 & 20.3 & \multicolumn{1}{c|}{16.9} & 31.9 \\
		AvT~\cite{prasad2021investigation} & None & {Noisy} & 55.1 & 39.5 & 31.1 & \multicolumn{1}{c|}{24.6} & 43.3 & 33.4 & 25.2 & \multicolumn{1}{c|}{20.9} & 40.8 & 33.4 & 29.3 & \multicolumn{1}{c|}{23.2} & 32.0 & 26.3 & 21.4 & \multicolumn{1}{c|}{18.5} & 31.1 \\
		\hline
		\multirow{4}{*} 
        {Wav2vec2.0~\cite{baevski2020wav2vec}} & None & Clean & 93.1 & 84.9 & 73.7 & \multicolumn{1}{c|}{58.0} & 80.6 & 72.7 & 59.7 & \multicolumn{1}{c|}{48.8} & 79.6 & 69.6 & 56.5 & \multicolumn{1}{c|}{42.4} & 68.9 & 58.1 & 43.7 & \multicolumn{1}{c|}{34.8} & 64.1 \\
		& None & {Noisy} & 70.8 & 58.6 & 47.6 & \multicolumn{1}{c|}{39.8} & 59.5 & 49.4 & 40.6 & \multicolumn{1}{c|}{35.4} & 56.5 & 48.4 & 42.4 & \multicolumn{1}{c|}{35.5} & 49.6 & 42.7 & 34.6 & \multicolumn{1}{c|}{31.8} & 46.5 \\
	& Clean & {Noisy} & 58.3 & 45.0 & 35.1 &   
        \multicolumn{1}{c|}{27.6} & 48.1 & 37.3 & 28.8 & \multicolumn{1}{c|}{24.7} & 44.9 & 36.1 & 29.2 &\multicolumn{1}{c|}{24.5} & 36.4 & 29.1 & 23.9 & \multicolumn{1}{c|}{19.2} & 34.3 \\
		& {Noisy} & {Noisy} & 47.4 & 35.9 & 29.0 & \multicolumn{1}{c|}{23.5} & 39.6 & 29.4 & 23.7 & \multicolumn{1}{c|}{20.2} & 40.9 & 32.4 & 26.8 & \multicolumn{1}{c|}{21.3} & 28.0 & 24.6 & 19.7 & \multicolumn{1}{c|}{17.0} & 28.7 \\
		\hline
		EW2~\cite{zhu2022noise} & {Noisy} & {Noisy} & 41.0 & 30.2 & 25.1 & \multicolumn{1}{c|}{19.2} & 33.4 & 24.4 & 19.7 & \multicolumn{1}{c|}{16.9} & 32.4 & 25.6 & 21.5 & \multicolumn{1}{c|}{17.6} & 24.7 & 21.4 & 17.7 & \multicolumn{1}{c|}{15.1} & 24.1 \\
		\hline
		\multirow{2}{*}{EW2 + SEW2~\cite{zhu2022joint}} & None & {Noisy} & 53.9 & 41.0 & 31.4 & \multicolumn{1}{c|}{27.4} & 43.0 & 32.4 & 26.8 & \multicolumn{1}{c|}{23.4} & 42.8 & 34.3 & 28.5 & \multicolumn{1}{c|}{22.9} & 32.8 & 27.6 & 24.0 & \multicolumn{1}{c|}{19.9} & 32.0 \\
		& {Noisy} & {Noisy} & 37.8 & 28.3 & 23.9 & \multicolumn{1}{c|}{18.6} & 31.5 & 23.2 & 18.8 & \multicolumn{1}{c|}{16.3} & 31.2 & 24.6 & 20.8 & \multicolumn{1}{c|}{17.2} & 23.5 & 20.6 & 17.2 & \multicolumn{1}{c|}{14.9} & 23.0 \\
		\hline
		\multirow{2}{*}{Wav2code (ours)} & None & {Noisy} & 51.6 & 38.9 & 29.7 & \multicolumn{1}{c|}{26.0} & 41.6 & 31.4 & 25.5 & \multicolumn{1}{c|}{22.5} & 40.7 & 32.4 & 26.9 & \multicolumn{1}{c|}{21.7} & 31.3 & 26.2 & 22.9 & \multicolumn{1}{c|}{19.4} & 30.5 \\
		& {Noisy} & {Noisy} & \textbf{36.3} & \textbf{27.2} & \textbf{23.3} & \multicolumn{1}{c|}{\textbf{18.2}} & \textbf{30.2} & \textbf{22.4} & \textbf{18.3} & \multicolumn{1}{c|}{\textbf{16.1}} & \textbf{30.3} & \textbf{24.0} & \textbf{20.4} & \multicolumn{1}{c|}{\textbf{16.9}} & \textbf{22.7} & \textbf{20.0} & \textbf{16.7} & \multicolumn{1}{c|}{\textbf{14.6}} & \textbf{22.5} \\
		\hline
	\end{tabular}
}
\vspace{-0.2cm}
\end{table*}

\begin{table}[t]
\caption{The performance of different approaches on type-B noisy test sets when EW2 backbone is pre-trained on LibriSpeech 460h data and Wav2code is pre-trained and finetuned on LibriSpeech 100h data.
``Avg'' denotes the averaged WER results on all type-B noise conditions.
}
\label{table3}
\centering
\resizebox{1.0\linewidth}{!}{
\begin{tabular}{l|cccccc}
    \hline
    \multirow{3}{*}{\textbf{Method}} & \multicolumn{6}{c}{\textbf{WER under SNR (dB)}} \\ \cline{2-7} 
    & \multicolumn{1}{c|}{\textbf{Babble}} & \multicolumn{1}{c|}{\textbf{Airport/Station}} & \multicolumn{1}{c|}{\textbf{AC/Vacuum}} & \multicolumn{1}{c|}{\textbf{Cafe}} & \multicolumn{1}{c|}{\textbf{Avg}} & \multirow{2}{*}{\textbf{Clean}} \\ \cline{2-6} 
    & \multicolumn{1}{c|}{\textbf{0$\sim$20 dB}} & \multicolumn{1}{c|}{\textbf{0$\sim$20 dB}} & \multicolumn{1}{c|}{\textbf{0$\sim$20 dB}} & \multicolumn{1}{c|}{\textbf{0$\sim$20 dB}} & \multicolumn{1}{c|}{\textbf{0$\sim$20 dB}} & \\ 
    \hline
    EW2~\cite{zhu2022noise} & \multicolumn{1}{c|}{21.0} & \multicolumn{1}{c|}{16.1} & \multicolumn{1}{c|}{15.7} & \multicolumn{1}{c|}{12.2} & \multicolumn{1}{c|}{16.3} & {7.5} \\
    EW2 + SEW2~\cite{zhu2022joint} & \multicolumn{1}{c|}{18.6} & \multicolumn{1}{c|}{14.4} & \multicolumn{1}{c|}{14.1} & \multicolumn{1}{c|}{11.2} & \multicolumn{1}{c|}{14.6} & {5.7} \\ \hline
    Wav2code (ours) & \multicolumn{1}{c|}{\textbf{17.3}} & \multicolumn{1}{c|}{\textbf{13.4}} & \multicolumn{1}{c|}{\textbf{13.2}} & \multicolumn{1}{c|}{\textbf{10.6}} & \multicolumn{1}{c|}{\textbf{13.6}} & {\textbf{5.3}} \\
    \hline
\end{tabular}
}
\vspace{-0.2cm}
\end{table}

\section{Results and Analysis}
\label{sec:results_analysis}
In this section, we present experimental results and analysis to verify the effectiveness of our proposed approach.

\subsection{Evaluation of the proposed Wav2code}
\label{ssec:eval_wav2code}
\textbf{Baselines:} Work~\cite{prasad2021investigation} provides a baseline on LibriSpeech-FreeSound benchmark, with Deepspeech2 model~\cite{amodei2016deep} and CTC objective for system training.
It also introduces DEMUCS speech enhancement module~\cite{defossez2020real} and gradient reversal layer (AvT) to improve noise robustness.
Wav2vec2.0~\cite{baevski2020wav2vec} leverages self-supervised learning to improve ASR performance, and EW2~\cite{zhu2022noise} build an enhanced version of it to further strengthen robustness.
Furthermore, latest work EW2 + SEW2~\cite{zhu2022joint} combines speech enhancement (SE) and self-supervised learning (SSL) to achieve the state-of-the-art.

Table~\ref{table1} and~\ref{table2} compare the ASR performance between our proposed Wav2code and the baselines in terms of WER, on type-A (stationary) and type-B (non-stationary) testing noises respectively.
Benchmark work~\cite{prasad2021investigation} find that introducing DEMUCS-based speech enhancement as front-end can significantly improve the performance of noisy ASR.
Then, Wav2vec2.0~\cite{baevski2020wav2vec} with clean data finetuning is less optimal and only comparable to the earliest Baseline in~\cite{prasad2021investigation}, and finetuning it on noisy data can significantly improve the testing results on most noisy testing conditions, \textit{i.e.}, stronger noise robustness, even though the performance on clean test set degrades.
Furthermore, introducing pre-training stage is beneficial to ASR performance on all test sets, where leveraging noisy data for pre-training can yield more improvements.
The best performance of Wav2vec2.0 outperforms that in baseline work~\cite{prasad2021investigation} with the power of SSL pre-training.
Based on that, EW2~\cite{zhu2022noise} achieves stronger robustness by performing noisy-to-clean reconstruction in self-supervised pre-training, and EW2 + SEW2~\cite{zhu2022joint} further introduces a joint training approach of SE and SSL to benefit from the effectiveness of both, which has achieved the state-of-the-art noise robustness on LibriSpeech-FreeSound benchmark.

Though effective, EW2 + SEW2 still suffers from the speech distortion problem caused by conventional SE, and it also fails to fully utilize the clean training data.
Our proposed Wav2code framework makes full use of unlabeled clean speech data as prior to implement {a feature-level SE with reduced} distortions for downstream ASR, which achieves the state-of-the-arts under all noisy testing conditions, and the performance on clean test set is also improved.
Furthermore, to verify that the effectiveness of Wav2code is not due to EW2 backbone pre-training, we also compares it with the EW2 + SEW2 under non-pretraining setting, where consistent improvements under all testing conditions demonstrate the origin effectiveness of our proposed Wav2code.
Meanwhile, it also indicates that EW2 pre-training contributes to the final superior performance of Wav2code.
Moreover, we also evaluate the Wav2code with larger-data pre-trained EW2 backbone, \textit{i.e.}, LibriSpeech 460-hour data.
As illustrated in Table~\ref{table3}, the proposed Wav2code also outperforms previous state-of-the-art under various noisy conditions, which verifies its strong generality.

\subsection{Evaluation of Codebook and Code Predictor}
\label{ssec:eval_codebook_codepred}
\begin{figure}[t]
  \centering
  \includegraphics[width=0.48\textwidth]{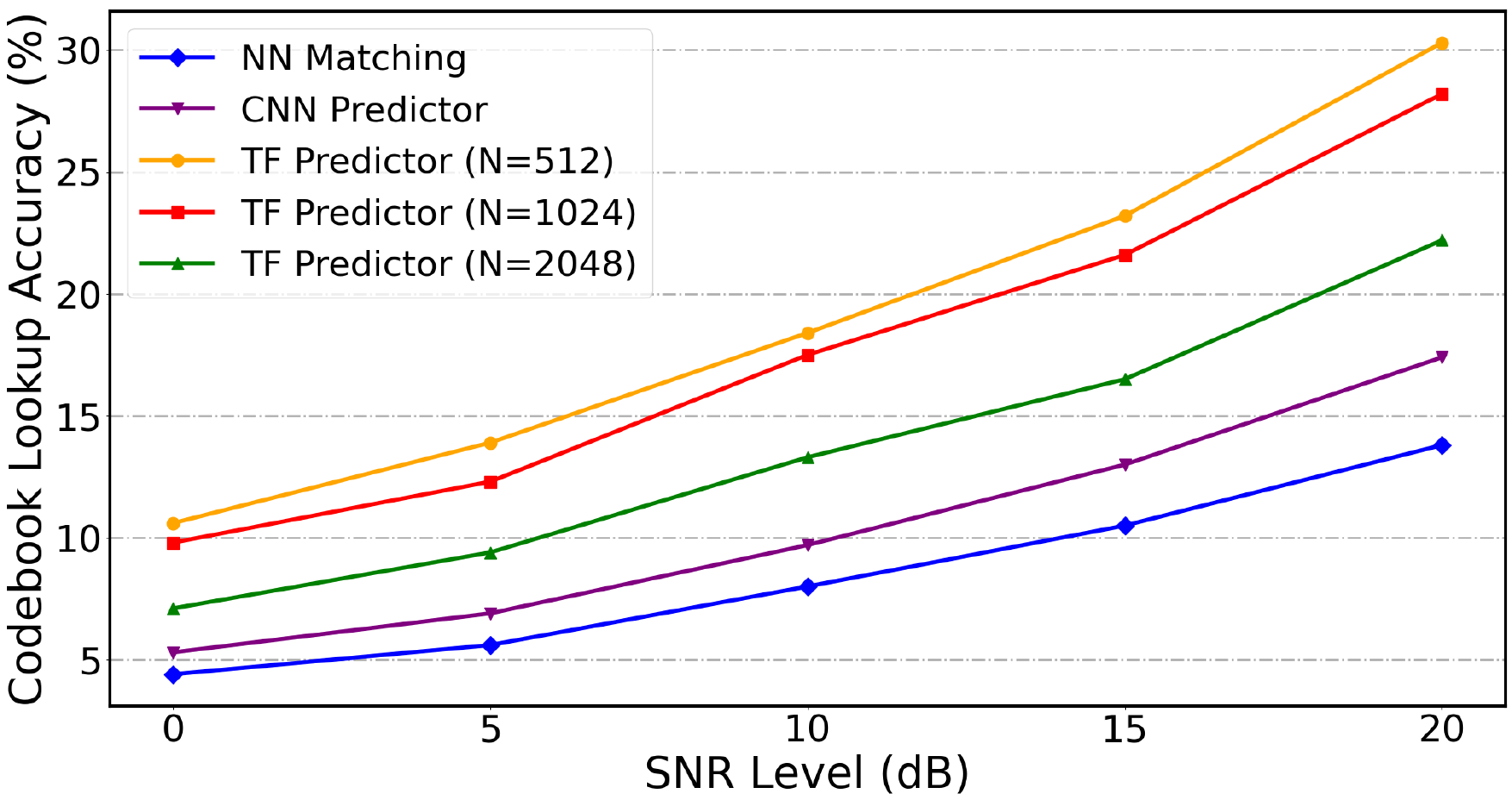}
  \caption{Comparison on code prediction accuracy under different SNR levels. 
  The results have been averaged over all FreeSound noise types.}
  \label{fig4}
  \vspace{-0.2cm}
\end{figure}

Table~\ref{table4} presents the ablation study on codebook and code predictor to verify their effectiveness in Wav2code framework.
We first investigate the importance of codebook space.
As shown in Exp. (1) of Table~\ref{table4}, removing the codebook in Wav2code (\textit{i.e.}, directly finetuning same as EW2 baseline) results in worse ASR performance under all testing conditions, which indicates the key role of discrete space of codebook to the effectiveness of Wav2code.
Furthermore, the effect of number of codebook entries $N$ is analyzed in Table~\ref{table5}.
Comparison between Exp. (7)-(9) suggests that $1024$ codebook entries is the most appropriate under our setting, where $N=512$ may not be sufficient to restore clean representations, and $N=2048$ only yields limited improvement as the redundancy in codebook entries could cause ambiguous code prediction in finetuning stage (see \textit{green line} in Fig.~\ref{fig4}).

\begin{table*}[t]
\caption{Ablation studies of network components and codebook lookup methods on LibriSpeech 100-hour setting.
No ``Codebook'' means the Wav2code framework degenerates into EW2 baseline, ``PT Codebook'' denotes pre-train codebook (Wav2code pre-training stage), and ``Fix Codebook'' denotes fix codebook during finetuning.
``NN'' denotes nearest-neighbor matching, ``TF'' denotes Transformer.
The Exp. ID in bold denotes our employed configurations.}
\label{table4}
\centering
\resizebox{1.0\textwidth}{!}{
\begin{tabular}{c|ccc|ccc|cccccccc}
    \hline
    \multirow{2}{*}{\textbf{Exp.}} & \multicolumn{3}{c|}{\textbf{Network}} & \multicolumn{3}{c|}{\textbf{Codebook Lookup}} & \multicolumn{8}{c}{\textbf{WER}} \\ \cline{2-15} 
    & \textbf{Codebook} & \textbf{PT Codebook} & \textbf{Fix Codebook} & \textbf{NN} & \textbf{CNN Predictor} & \textbf{TF Predictor} & \multicolumn{1}{c|}{\textbf{Traffic}} & \multicolumn{1}{c|}{\textbf{Metro}} & \multicolumn{1}{c|}{\textbf{Car}} & \multicolumn{1}{c|}{\textbf{Babble}} & \multicolumn{1}{c|}{\textbf{\begin{tabular}[c]{@{}c@{}}Airport/\\ Station\end{tabular}}} & \multicolumn{1}{c|}{\textbf{\begin{tabular}[c]{@{}c@{}}AC/\\ Vacuum\end{tabular}}} & \multicolumn{1}{c|}{\textbf{Cafe}} & \textbf{Avg} \\
    \hline
    1 & & & & & & & \multicolumn{1}{c|}{20.9} & \multicolumn{1}{c|}{19.8} & \multicolumn{1}{c|}{14.9} & \multicolumn{1}{c|}{33.9} & \multicolumn{1}{c|}{27.4} & \multicolumn{1}{c|}{27.9} & \multicolumn{1}{c|}{22.1} & 23.8 \\
    2 & \cmark & \cmark & \cmark & \cmark & & & \multicolumn{1}{c|}{20.7} & \multicolumn{1}{c|}{19.6} & \multicolumn{1}{c|}{14.8} & \multicolumn{1}{c|}{33.4} & \multicolumn{1}{c|}{27.1} & \multicolumn{1}{c|}{27.8} & \multicolumn{1}{c|}{21.8} & 23.6 \\
    3 & \cmark & \cmark & \cmark & & \cmark & & \multicolumn{1}{c|}{20.1} & \multicolumn{1}{c|}{19.3} & \multicolumn{1}{c|}{14.5} & \multicolumn{1}{c|}{32.5} & \multicolumn{1}{c|}{26.7} & \multicolumn{1}{c|}{27.5} & \multicolumn{1}{c|}{21.4} & 23.1 \\
    4 & \cmark & \cmark & & & & \cmark & \multicolumn{1}{c|}{19.6} & \multicolumn{1}{c|}{19.0} & \multicolumn{1}{c|}{13.9} & \multicolumn{1}{c|}{31.9} & \multicolumn{1}{c|}{26.0} & \multicolumn{1}{c|}{26.6} & \multicolumn{1}{c|}{20.8} & 22.5 \\
    5 & \cmark & & & & & \cmark & \multicolumn{1}{c|}{20.3} & \multicolumn{1}{c|}{19.4} & \multicolumn{1}{c|}{14.6} & \multicolumn{1}{c|}{33.1} & \multicolumn{1}{c|}{26.8} & \multicolumn{1}{c|}{27.6} & \multicolumn{1}{c|}{21.5} & 23.3 \\
    \hline
    \textbf{6 (ours)} & \cmark & \cmark & \cmark & & & \cmark & \multicolumn{1}{c|}{\textbf{18.9}} & \multicolumn{1}{c|}{\textbf{18.4}} & \multicolumn{1}{c|}{\textbf{13.5}} & \multicolumn{1}{c|}{\textbf{30.5}} & \multicolumn{1}{c|}{\textbf{24.9}} & \multicolumn{1}{c|}{\textbf{26.1}} & \multicolumn{1}{c|}{\textbf{20.6}} & \textbf{21.8}  \\
    \hline
\end{tabular}
}
\vspace{-0.2cm}
\end{table*}

\begin{table}[t]
\caption{Ablation studies of the configurations in codebook and code predictor on LibriSpeech 100-hour setting.
The hyper-parameter $N$ denotes number of codebook entries, and $M$ denotes the number of Transformer blocks in code predictor.
The Exp. ID in bold denotes our used configurations.}
\label{table5}
\centering
\resizebox{1.0\linewidth}{!}{
	\begin{tabular}{c|cc|cccccccc}
		\hline
		\multirow{2}{*}{\textbf{Exp.}} & \multicolumn{2}{c|}{\textbf{Config}} & \multicolumn{8}{c}{\textbf{WER}} \\ \cline{2-11} 
	    & $\bm{N}$ & $\bm{M}$ & \multicolumn{1}{c|}{\textbf{Traffic}} & \multicolumn{1}{c|}{\textbf{Metro}} & \multicolumn{1}{c|}{\textbf{Car}} & \multicolumn{1}{c|}{\textbf{Babble}} & \multicolumn{1}{c|}{\textbf{\begin{tabular}[c]{@{}c@{}}Airport/\\ Station\end{tabular}}} & \multicolumn{1}{c|}{\textbf{\begin{tabular}[c]{@{}c@{}}AC/\\ Vacuum\end{tabular}}} & \multicolumn{1}{c|}{\textbf{Cafe}} & \textbf{Avg} \\
		\hline
		7 & 512 & 4 & \multicolumn{1}{c|}{19.8} & \multicolumn{1}{c|}{19.3} & \multicolumn{1}{c|}{14.2} & \multicolumn{1}{c|}{31.9} & \multicolumn{1}{c|}{25.9} & \multicolumn{1}{c|}{26.8} & \multicolumn{1}{c|}{21.1} & 22.7  \\
		\textbf{8} & 1024 & 4 & \multicolumn{1}{c|}{18.9} & \multicolumn{1}{c|}{18.4} & \multicolumn{1}{c|}{13.5} & \multicolumn{1}{c|}{30.5} & \multicolumn{1}{c|}{24.9} & \multicolumn{1}{c|}{26.1} & \multicolumn{1}{c|}{20.6} & 21.8 \\
		9 & 2048 & 4 & \multicolumn{1}{c|}{18.7} & \multicolumn{1}{c|}{18.2} & \multicolumn{1}{c|}{13.2} & \multicolumn{1}{c|}{30.1} & \multicolumn{1}{c|}{24.6} & \multicolumn{1}{c|}{25.9} & \multicolumn{1}{c|}{20.3} & 21.6  \\
		\hline
		10 & 1024 & 2 & \multicolumn{1}{c|}{19.3} & \multicolumn{1}{c|}{18.9} & \multicolumn{1}{c|}{13.8} & \multicolumn{1}{c|}{31.1} & \multicolumn{1}{c|}{25.3} & \multicolumn{1}{c|}{26.4} & \multicolumn{1}{c|}{20.8} & 22.2  \\
		\textbf{11} & 1024 & 4 & \multicolumn{1}{c|}{18.9} & \multicolumn{1}{c|}{18.4} & \multicolumn{1}{c|}{13.5} & \multicolumn{1}{c|}{30.5} & \multicolumn{1}{c|}{24.9} & \multicolumn{1}{c|}{26.1} & \multicolumn{1}{c|}{20.6} & 21.8 \\
		12 & 1024 & 8 & \multicolumn{1}{c|}{18.8} & \multicolumn{1}{c|}{18.3} & \multicolumn{1}{c|}{13.4} & \multicolumn{1}{c|}{30.4} & \multicolumn{1}{c|}{24.7} & \multicolumn{1}{c|}{26.0} & \multicolumn{1}{c|}{20.4} & 21.7  \\
		\hline
	\end{tabular}
}
\vspace{-0.2cm}
\end{table}

To further verify the effectiveness of our Transformer-based code predictor for codebook lookup, we compare it with two baseline schemes, \textit{i.e.}, Nearest-neighbor (NN) matching in Exp. (2) and a two-layer CNN code predictor in Exp. (3).
As shown in Table~\ref{table4}, comparison between Exp. (2) and (3) suggests that code predictor is more effective than NN matching, as the latter is infeasible for codebook lookup due to the corruption in input noisy representations (see Fig.~\ref{fig5}(b)).
However, the local nature of convolution in CNN restricts its capacity of long sequence code prediction.
In comparison, our Transformer-based code predictor yields better ASR performance thanks to its stronger ability of global context modeling, \textit{i.e.}, Exp. (6) vs. (3).
As shown in Fig.~\ref{fig4}, the Transformer code predictor achieves higher prediction accuracy than CNN predictor and NN matching.
Furthermore, comparison between Exp. (6) and (4) indicates the importance of fixing codebook during finetuning to protect the pre-trained clean speech prior.
Comparison between Exp. (6) and (5) demonstrates that the pre-trained prior knowledge in codebook is important to the clean speech restoration and downstream ASR.
In addition, Table~\ref{table5} presents the effect of Transformer block numbers $M$ in Exp. (10)-(12).
We observe that employing $4$ Transformer blocks achieves considerable improvements over $2$ blocks, while further increasing it to $8$ can only result in limited gains, which means it could approach the upper-bound of Transformer code predictor under our experimental setting.
As a result, we select $M=4$ in all experiments without specified.

\subsection{Evaluation of Interactive Feature Fusion Network}
\label{ssec:eval_iffnet}
Table~\ref{table6} presents the ablation study on interactive feature fusion network (IFF-Net) to investigate its effectiveness in Wav2code.
As shown in Exp. (13), directly using restored speech representation for downstream ASR results in sub-optimal performance, as it may suffer from some loss of feature fidelity (see Fig.~\ref{fig7}(b)).
Combining original noisy and the restored speech representations by concatenation alleviates it and yields some improvements, \textit{i.e.}, Exp. (14). 
In comparison, our proposed IFF-Net achieves significantly better performance by interactively fuse the fidelity and quality in final representation, \textit{i.e.}, Exp. (18) vs. (14).
In addition, comparison between Exp. (15)-(18) reflects the effect of hyper-parameters on IFF-Net.
We observe that both the number of bottleneck modules $N_{rep}$ and the bottleneck dimension $D'$ contributes to the effect of IFF-Net.
The best performance is achieved with large $N_{rep}$ and $D'$, while further increasing them would not yield significant gains of performance.

\begin{table*}[t]
\caption{Ablation studies of fusion strategies on LibriSpeech 100-hour setting.
``Fusion Strategy'' denotes the strategy to fuse the noisy and restored speech representations, where ``None'' means using only restored speech representation for downstream ASR.
The hyper-parameter $N_{rep}$ denotes the repeated times of bottleneck modules in IFF-Net, and $D'$ denotes its embedding dimension.}
\label{table6}
\centering
\resizebox{0.9\textwidth}{!}{
	\begin{tabular}{c|ccc|cc|cccccccc}
		\hline
		\multirow{2}{*}{\textbf{Exp.}} & \multicolumn{3}{c|}{\textbf{Fusion Strategy}} & \multicolumn{2}{c|}{\textbf{Config}} & \multicolumn{8}{c}{\textbf{WER}} \\ \cline{2-14} 
	    & \textbf{None} & \textbf{Concat} & \textbf{IFF-Net} & $\bm{N_{rep}}$ & $\bm{D'}$ & \multicolumn{1}{c|}{\textbf{Traffic}} & \multicolumn{1}{c|}{\textbf{Metro}} & \multicolumn{1}{c|}{\textbf{Car}} & \multicolumn{1}{c|}{\textbf{Babble}} & \multicolumn{1}{c|}{\textbf{\begin{tabular}[c]{@{}c@{}}Airport/\\ Station\end{tabular}}} & \multicolumn{1}{c|}{\textbf{\begin{tabular}[c]{@{}c@{}}AC/\\ Vacuum\end{tabular}}} & \multicolumn{1}{c|}{\textbf{Cafe}} & \textbf{Avg} \\
		\hline
		13 & \cmark & & & & & \multicolumn{1}{c|}{20.4} & \multicolumn{1}{c|}{19.4} & \multicolumn{1}{c|}{14.6} & \multicolumn{1}{c|}{33.1} & \multicolumn{1}{c|}{26.8} & \multicolumn{1}{c|}{27.4} & \multicolumn{1}{c|}{21.6} & 23.3  \\
		14 & & \cmark & & & & \multicolumn{1}{c|}{20.1} & \multicolumn{1}{c|}{19.2} & \multicolumn{1}{c|}{14.4} & \multicolumn{1}{c|}{32.6} & \multicolumn{1}{c|}{26.4} & \multicolumn{1}{c|}{27.3} & \multicolumn{1}{c|}{21.4} & 23.1  \\
		15 & & & \cmark & 2 & 64 & \multicolumn{1}{c|}{19.6} & \multicolumn{1}{c|}{18.8} & \multicolumn{1}{c|}{14.1} & \multicolumn{1}{c|}{31.8} & \multicolumn{1}{c|}{25.7} & \multicolumn{1}{c|}{26.9} & \multicolumn{1}{c|}{21.1} & 22.6 \\
		16 & & & \cmark & 2 & 128 & \multicolumn{1}{c|}{19.3} & \multicolumn{1}{c|}{18.7} & \multicolumn{1}{c|}{13.9} & \multicolumn{1}{c|}{31.2} & \multicolumn{1}{c|}{25.5} & \multicolumn{1}{c|}{26.7} & \multicolumn{1}{c|}{20.8} & 22.3 \\
		17 & & & \cmark & 4 & 64 & \multicolumn{1}{c|}{19.1} & \multicolumn{1}{c|}{18.6} & \multicolumn{1}{c|}{13.8} & \multicolumn{1}{c|}{30.9} & \multicolumn{1}{c|}{25.2} & \multicolumn{1}{c|}{26.6} & \multicolumn{1}{c|}{20.5} & 22.1 \\
		\hline
		\textbf{18 (ours)} & & & \cmark & 4 & 128 & \multicolumn{1}{c|}{\textbf{18.9}} & \multicolumn{1}{c|}{\textbf{18.4}} & \multicolumn{1}{c|}{\textbf{13.5}} & \multicolumn{1}{c|}{\textbf{30.5}} & \multicolumn{1}{c|}{\textbf{24.9}} & \multicolumn{1}{c|}{\textbf{26.1}} & \multicolumn{1}{c|}{\textbf{20.6}} & \textbf{21.8} \\
		\hline
	\end{tabular}
}
\vspace{-0.1cm}
\end{table*}

\subsection{Visualizations of Codebook Entries}
\label{ssec:visual_codebook}

\begin{figure}[t]
  \centering
  \includegraphics[width=0.49\textwidth]{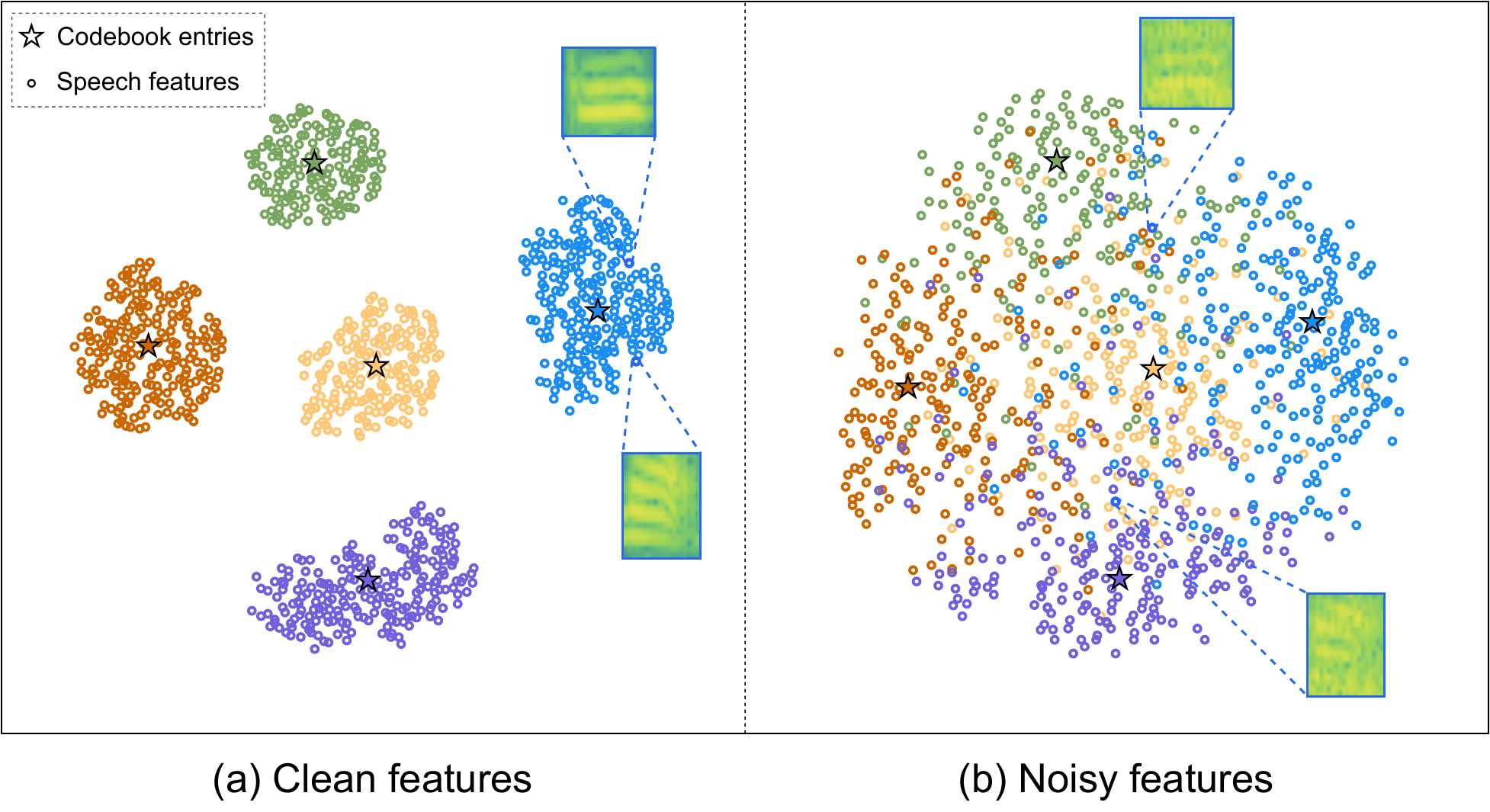}
  \caption{t-SNE visualizations of clean/noisy speech features and codebook entries.
  Features of same color are extracted from parallel clean/noisy speech data.
  Examples show that noise corruption increases the diversity and uncertainty of speech features, making it challenging for correct code assignment.
  }
  \label{fig5}
  \vspace{-0.2cm}
\end{figure}

In order to better visualize the codebook entries and speech features, we build an auto-encoder model on Wav2code to implement explicit speech enhancement.
The auto-encoder model consists a pair of encoder and decoder following prior work~\cite{defossez2022high}, where the code predictor, codebook and IFF-Net are all placed between them.
Encoder contains four strided convolution based down-sampling blocks, and decoder mirrors the encoder with four transposed convolution based up-sampling blocks, where the configuration details are same as~\cite{defossez2022high}.

Fig.~\ref{fig5} visualizes the clean/noisy speech features and the pre-trained codebook entries.
We can observe that the clean speech features are distributed compactly around the corresponding codebook entries.
However, the noise corruption deviates the features from the correct code and spreads them into nearby code clusters, leading to undesirable restoration results by near-neighbor matching as depicted in Fig.~\ref{fig6}(c).

\vspace{-0.2cm}
\subsection{Visualizations of Restored Clean Speech Representations}
\label{ssec:visual_restored}
\begin{figure*}[t]
  \centering
  \includegraphics[width=1.0\textwidth]{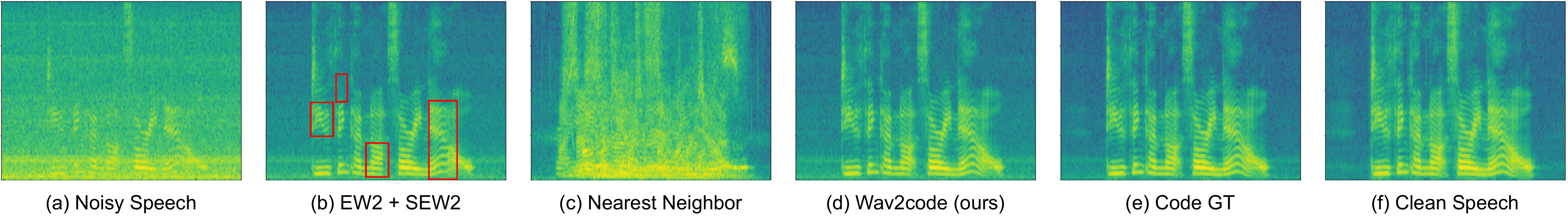}
  \vspace{-0.5cm}
  \caption{Visualizations of speech features.
  (a) Noisy speech.
  (b) Enhanced speech from EW2 + SEW2~\cite{zhu2022joint}, where speech distortions are highlighted in red boxes.
  (c) Results of codebook prior by Nearest-Neighbor (NN) matching, where we can observe serious speech distortions caused by wrong code assignments.
  (d) Restored speech ({with reduced} distortions) from our proposed Wav2code.
  (e) Reconstruction results from the code sequence ground truth.
  (f) Clean speech.}
  \label{fig6}
  \vspace{-0.2cm}
\end{figure*}

Fig.~\ref{fig6} presents the spectrograms of clean/noisy speech, enhanced speech from baseline and the restored clean speech from our proposed Wav2code.
First, comparison between (a) and (f) indicates the significant noise corruption in noisy speech.
The enhanced speech from baseline EW2 + SEW2~\cite{zhu2022joint} effectively reduce the noise and improves the speech quality as shown in (b), but there exists some speech distortions (red boxes) compared to the ground-truth clean speech.
Then in our experiments, nearest-neighbor matching fails to restore desired clean speech, as the noise corruption deviates the speech features from the correct codebook entry as presented in Fig.~\ref{fig5}(b).
In comparison, our proposed Wav2code effectively restores high-quality clean speech {with reduced} distortions thanks to the pre-trained clean prior as well as global sequence modeling in code prediction, which moves one step forward to the ground-truth clean features.
Fig.~\ref{fig6}(e) presents the reconstruction results from the ground-truth code sequence, which effectively reconstruct the clean speech.
Therefore, it shows the strong expressiveness of codebook space, which can perceptually approximates the clean speech space and thus restore high-quality clean representations.

\begin{figure}[t]
  \centering
  \includegraphics[width=0.5\textwidth]{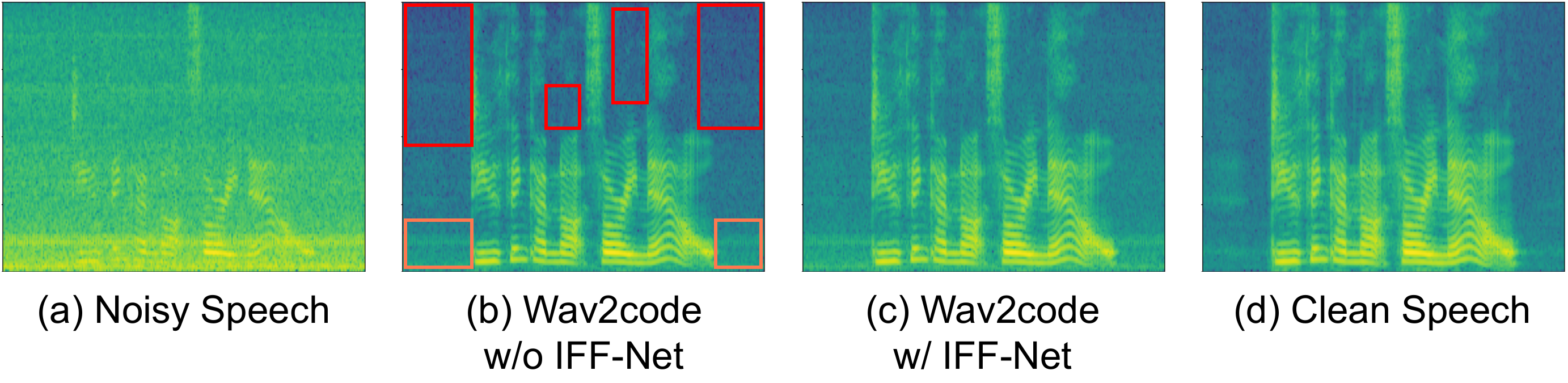}
  \vspace{-0.5cm}
  \caption{Effect of IFF-Net on combining quality and fidelity of speech.
  (a) Noisy speech.
  (b) Restored speech from Wav2code without IFF-Net (good quality but less-optimal fidelity).
  (c) Restored speech from Wav2code with IFF-Net (good quality and fidelity).
  (d) Clean speech.
  Red and orange boxes highlight the loss of fidelity in restored speech from Wav2code without IFF-Net, \textit{i.e.}, uneven noise in high- and low-frequency bands.}
  \label{fig7}
  \vspace{-0.3cm}
\end{figure}

Furthermore, Fig.~\ref{fig7} illustrates the effect of proposed IFF-Net in combining the quality and fidelity of restored speech.
Fig.~\ref{fig7}(b) depicts the restored speech from codebook in Wav2code, which has effectively restored the clean speech {with reduced} distortions.
However, we can also observe some loss of fidelity, \textit{i.e.}, uneven noise distribution in high- and low-frequency bands.
As depicted by the red boxes, there is less high-frequency noise compared to ground-truth clean speech, but the low-frequency noise (orange boxes) still exists similar to noisy speech.
Such loss of restoration fidelity may explain why downstream ASR performance is sub-optimal when directly using the restored speech from codebook, as presented in Table~\ref{table6}.
After introducing IFF-Net to fuse original noisy and the restored speech, our Wav2code effectively restores the clean speech with both high fidelity and good quality, which thus results in better performance for downstream ASR (see Table~\ref{table6}).
As shown in Fig.~\ref{fig7}(c), the fused speech becomes significantly closer to ground-truth clean speech than the directly restored speech from codebook.

\subsection{Performance Evaluation on CHiME-4 Dataset}
\label{ssec:performance_chime4}
Finally, we evaluate the proposed Wav2code in real noisy conditions with CHiME-4 dataset.
Table~\ref{table7} presents the performance comparison on CHiME-4 real data, \textit{i.e.}, validation set `dt05\_real' and test set `et05\_real'.
In terms of various baselines on CHiME4 data, we split them into two categories, \textit{i.e.}, supervised and self-supervised methods.
The top part of Table~\ref{table7} show some representative supervised methods, which focus on data augmentation and processing techniques, \textit{e.g.}, speaker adaptation.
Then, the middle part presents the recent popular self-supervised methods, which benefits from large amount of unlabeled data during pre-training.
For example, the popular supervised method Wav2vec2.0 Base~\cite{baevski2020wav2vec} achieves WERs of 10.5/17.3 on dt05\_real/et05\_real sets, and it achieves lower WERs of 3.8/7.5 when using Transformer LM for rescoring during inference, which is similar to other self-supervised methods like HuBERT~\cite{hsu2021hubert} and Data2vec~\cite{baevski2022data2vec}.
Based on that, many variants of Wav2vec2.0 has proposed to improve the noise robustness, where the Enhanced Wav2vec2.0 (EW2)~\cite{zhu2022noise} achieves the best performance with WERs of 9.4/15.6 on dt05\_real/et05\_real sets, as well as 3.5/5.9 with Transformer LM.
Furthermore, recent work EW2 + SEW2~\cite{zhu2022joint} propose to jointly train speech enhancement (SE) and self-supervised learning (SSL) to improve noise robustness, which has achieved significant improvement over EW2 with WERs of 8.2/14.3, as well as 3.0/5.9 with Transformer LM.
However, their method still suffers from the speech distortion problem (see Fig.~\ref{fig6}(b)).
To this end, our proposed Wav2code leverages codebook prior to restore clean speech {with reduced} distortions (see Fig.~\ref{fig6}(d)), which results in better performance than EW2 + SEW2 baseline, \textit{i.e.}, WERs of 7.6/13.7 without LM and 2.7/5.5 with Transformer LM.
In addition, we note that Chang et al.~\cite{chang2022end} achieves the state-of-the-art on CHiME-4 benchmark also with joint SE and SSL approach, { which could be attributed to their used pre-trained WavLM-Large front-end}.
To summarize, our Wav2code achieves superior performance on popular CHiME-4 dataset with improved noise robustness.

\begin{table}[t]
\vspace{-0.2cm}
\caption{The performance comparison of different approaches on CHiME-4 dataset (one-channel track).
}
\label{table7}
\centering
\begin{tabular}{lccc}
\hline
\multirow{2}{*}{\textbf{Model}} & \multirow{2}{*}{\textbf{LM}} & \multicolumn{2}{c}{\textbf{WER}} \\ \cline{3-4}
& & \textbf{dt05\_real} & \textbf{et05\_real} \\ \hline
\hline
\multicolumn{4}{l}{\textbf{Supervised}} \\
DNN baseline~\cite{vincent20164th} & N-gram & 11.6 & 23.7 \\
Du et al.~\cite{du2016ustc} & LSTM & 4.5 & 9.2 \\
Menne et al.~\cite{menne2016rwth} & LSTM & 5.1 & 9.3 \\
Wang et al.~\cite{wang2020complex} & LSTM & 3.5 & 6.8 \\ \hline
\hline
\multicolumn{4}{l}{\textbf{Self-supervised}} \\
Wang et al. (960h)~\cite{wang2022improving} & LSTM & 5.0 & 9.0 \\
Wang et al. (60kh)~\cite{wang2022improving} & LSTM & 2.8 & 5.8 \\
\hline
Wav2vec2.0 Base~\cite{gao2021data} & None & 10.3 & 17.8 \\
Gao et al.~\cite{gao2021data} & None & 8.7 & 15.8 \\
\hline
\multirow{2}{*}{HuBERT Base~\cite{hsu2021hubert}} & None & 10.4 & 17.0 \\
& LSTM & 3.8 & 7.1 \\
\hline
\multirow{2}{*}{Data2vec Base~\cite{zhu2022robust}} & None & 9.5 & 15.7 \\
& Transformer & 3.5 & 6.5 \\
\multirow{2}{*}{Robust data2vec~\cite{zhu2022robust}} & None & 8.3 & 12.8 \\
& Transformer & 3.1 & 5.8 \\
\hline
\multirow{2}{*}{Wav2vec2.0 Base~\cite{wang2022wav2vec}} & None & 10.6 & 17.6 \\
& LSTM & 3.7 & 7.2 \\
\multirow{2}{*}{Wav2vec-switch~\cite{wang2022wav2vec}} & None & 10.0 & 16.5 \\
& LSTM & 3.5 & 6.6 \\
\hline
\multirow{2}{*}{Wav2vec2.0 Base~\cite{zhu2022noise}} & None & 10.5 & 17.3 \\
& Transformer & 3.8 & 7.5 \\
\multirow{2}{*}{EW2~\cite{zhu2022noise}} & None & 9.4 & 15.6 \\
& Transformer & 3.5 & 6.4 \\ \hline
\multicolumn{4}{l}{\textbf{Speech Enhancement + Self-supervised}}\\
Chang et al.~\cite{chang2022end} & Transformer & 2.03 & 3.92 \\
\hline
\multirow{2}{*}{EW2 + SEW2~\cite{zhu2022joint}} & None & 8.2 & 14.3 \\
& Transformer & 3.0 & 5.9 \\ \hline
\multirow{2}{*}{Wav2code (ours)} & None & 7.6 & 13.7 \\
& Transformer & 2.7 & 5.5 \\ \cline{2-4}
\hline
\end{tabular}
\vspace{-0.15cm}
\end{table}

\begin{table}[t]

\caption{Comparison of floating point operations (FLOPs) of different approaches.}
\label{table8}
\centering
\resizebox{0.7\linewidth}{!}{
    \begin{tabular}{c|cc}
    \hline
    FLOPs (Billion) & Pre-training & Finetuning \\ \hline
    Wav2vec 2.0~\cite{baevski2020wav2vec} & 20.3 & 20.3 \\
    EW2~\cite{zhu2022noise} & 20.3 & 20.3 \\
    EW2 + SEW2~\cite{zhu2022joint} & 23.6 & 23.6 \\ \hline
    Wav2code (ours) & 20.3 & 21.3 \\
    \hline   
    \end{tabular}
    \vspace{-0.2cm}
}
\end{table}

{
\subsection{Comparison of Computation Cost}
Table~\ref{table8} compares the computation cost in FLOPs between Wav2code and baselines.
Our Wav2code pre-training does not introduce extra cost as NN matching is very efficient, while finetuning stage introduces 1.0 billion extra cost due to Transformer-based code prediction and IFF-Net fusion.
In addition, EW2+SEW2~\cite{zhu2022joint} consumes the most computations due to the introduced DEMUCS module for SE.}

\section{Conclusion}
\label{sec:conclusion}
In this paper, we propose a self-supervised framework named Wav2code to implement a {feature-level speech enhancement with reduced} distortions for noise-robust ASR.
First, during pre-training the clean speech representations from SSL model are employed to lookup a discrete codebook via nearest-neighbor feature matching, the resulted code sequence are then exploited to reconstruct the original clean representations, in order to store them as prior in codebook.
Second, in finetuning stage we propose a Transformer-based code predictor to accurately predict clean codes by modeling the global dependency of input noisy representations, which enables restoration of high-quality clean representations {with less} distortions.
Furthermore, we propose an interactive feature fusion network to combine original noisy and the restored clean representations to consider both fidelity and quality, resulting in more informative features for downstream ASR.
Experiments on both synthetic and real noisy datasets demonstrate that our Wav2code can solve the speech distortions and improve ASR performance under various noisy conditions.



\bibliographystyle{IEEEtran}
\bibliography{strings,refs}

\end{document}